\documentclass[11pt]{article}
\usepackage{amsmath,amssymb}
\begin{document}
\begin{flushright} SU-4252-788
\\
\end{flushright}         
\begin{center}
\vskip 3em
{\LARGE NON-LINEAR SIGMA MODEL ON THE FUZZY} 
\vskip 1em 
{\LARGE SUPERSPHERE}
\vskip 2em
{\large Se\c{c}kin K\"{u}rk\c{c}\"{u}o\v{g}lu
\footnote[1]{E-mail: skurkcuo@phy.syr.edu} 
\\[2em]}
\em{\oddsidemargin 0 mm
Department of Physics, Syracuse University,
Syracuse, NY 13244-1130, USA}\\
\end{center}
\vskip 1em
\begin{abstract}
In this note we develop fuzzy versions of the supersymmetric non-linear sigma model on the supersphere $S^{(2,2)}$.
In \cite{bal} Bott projectors have been used to obtain the fuzzy ${\mathbb C}P^1$ model.
Our approach utilizes the use of supersymmetric extensions of these projectors. Here we obtain these
(super)-projectors and quantize them in a fashion similar to the one given in \cite{bal}. We discuss the 
interpretation of 
the resulting model as a finite dimensional matrix model.
\end{abstract}

\newpage

\setcounter{footnote}{0}

\section{Introduction}

In past few years studies of field theories on non-commutative manifolds have been very fruitful. To construct such theories one 
usually starts with a continuum theory on a manifold ${\cal M}$ and replaces the commutative algebra ${\cal A}$ of functions on 
${\cal M}$ by a non-commutative algebra $A$ which preserves most of the symmetries of the continuum theory and which approximates the 
commutative algebra ${\cal A}$ and hence the continuum theory in the commutative limit. It is possible to realize a large class of
such non-commutative field theories as finite dimensional matrix models. Field theories on the non-commutative (fuzzy) sphere
$S_F^2$ and the fuzzy supersphere $S_F^{(2,2)}$ are two such examples. As non-commutative manifolds the former is based on the 
irreducible representations of the $su(2)$ Lie algebra, whereas the latter is described by the irreducible representations of the Lie 
superalgebra $osp(2,1)$. To date many studies on different and novel aspects of field theories on $S_F^2$ have been carried out 
\cite{bal, paper1, paper2, paper3, paper4, paper5}. 

Recently, ${\mathbb C}P^1$ model on the fuzzy sphere $S_F^2$ have been studied from several different points of view 
{\cite{bal, bal1, trg}.
In \cite{bal} the commutative theory have been reformulated by replacing the non-linear fields with a certain class of projectors 
called ``Bott Projectors". A discrete (fuzzy) version of these projectors are easily obtained and they have permitted the 
construction of a fuzzy ${\mathbb C}P^1$ model in a rather straightforward way.

In this paper we address the question of constructing a fuzzy supersymmetric non-linear sigma model on $S^{(2,2)}$.
For this purpose we obtain the supersymmetric extensions of the Bott projectors and quantize them in a 
similar manner as discussed in \cite{bal}. Using the quantized (super)-projectors and the already known description 
of $S^{(2,2)}$ in terms of the Lie superalgebras $osp(2,1)$ and $osp(2,2)$ and their associated Lie supergroups we construct 
the fuzzy supersymmetric non-linear sigma model on $S^{(2,2)}$. We interpret the resulting theory as a finite dimensional matrix model
and comment on its various physical properties. 
   
\section{${\mathbb C}P^1$ Sigma Model and Bott Projectors}

Non-linear sigma models are customarily defined in terms of a field that maps the world-sheet to the target manifold. In the case of 
the ${\mathbb C}P^1$ models both world-sheet and the target manifolds are $2$-spheres ($S^2$) and the field $\vec{n}$ maps the point 
$x$ of the world-sheet
\begin{equation} 
S^2 = \big \langle x = (x_1, x_2, x_3) \in {\mathbb R^3} \big| \, \, x_i x_i =1 \big \rangle
\end{equation} 
to a point on the target manifold
\begin{equation}
S^2 = \big \langle \vec{n}(x)=(n_1(x), n_2(x), n_3(x)) \in {\mathbb R^3} \big| \, \, n_a(x) n_a(x) =1  \big \rangle \,.
\end{equation}
As is well known these maps are classified in terms of an integer $\kappa$ called the winding number since the second homotopy class
$\pi_2(S^2) = {\mathbb Z}$.

An alternative formulation of ${\mathbb C}P^1$ model which happens to be more convenient for passage to fuzzy ${\mathbb C}P^1$ 
model have been considered in \cite{bal}. 
This formulation uses certain class of projectors, known as Bott projectors instead of the non-linear fields. 
At the topological sector $\kappa = 1$ the Bott projector can be expressed in terms of $\vec{n}$ as 
\begin{equation}
P(x) = \frac{1 + \vec{\tau} \cdot \vec{n}(x)}{2}
\label{eq:projector1} 
\end{equation}
where $\vec{\tau}$ are the Pauli matrices. $P(x)$ is a projector since
$P^2(x) = P(x)$ and $P^\dagger(x) =P(x)$. At the topological sector $\kappa$, Bott projector can be expressed by introducing the 
partial isometries\footnote{To be more precise the partial isometry ${\vartheta}_\kappa^\dagger$
in the algebra ${\cal A} = C^\infty(S^3) \otimes Mat_{2 \times 2} {\mathbb C}$ is the matrix $ \left ( \begin{array}{cc} 
{\bar z}_1^\kappa & {\bar z}_2^\kappa \\ 0& 0 \\ \end{array} \right )$. But for all practical calculations it is perfectly safe to 
call (\ref{eq:partial1}) as the partial isometry, thus we do so from now on.} ${\vartheta}_\kappa^\dagger$ (for $\kappa > 0$) \cite{wegge-olsen} 
\begin{equation}
\vartheta_\kappa^\dagger (z) =
\left ( 
\begin{array}{cc}
{\bar z}_1^\kappa & {\bar z}_2^\kappa \\
\end{array}
\right ) \frac{1}{\sqrt{Z_\kappa}} \,, \quad
\vartheta_\kappa (z) =
\left ( 
\begin{array}{c}
 z_1^\kappa \\
 z_2^\kappa \\
\end{array}
\right ) \frac{1}{\sqrt{Z_\kappa}} \,,
\quad \quad Z_\kappa = |z_1|^{2 \kappa} +|z_2|^{2 \kappa}
\label{eq:partial1}
\end{equation}
where $z = (z_1, z_2)$ is a point on $S^3 = \langle z = (z_1, z_2) \in {\mathbb C}^2 \big | \, \, |z|^2 := |z_1|^2 + |z_2|^2 =1 
\rangle$ and ``bar'' stands for complex conjugation. Using the Hopf fibration $U(1) \rightarrow S^3 \rightarrow S^2$, points $x$ 
on the world-sheet $S^2$ is expressed in terms of $z$ as 
\begin{equation}
x_i = z^\dagger \tau_i z \,.
\label{eq:Hopff}
\end{equation}
By definition $\vartheta_\kappa^\dagger$ is a partial isometry if and only if $\vartheta_\kappa(z) \vartheta_\kappa^\dagger(z)$ is 
a projection. It is straightforward to check that $P_\kappa(x)$ in the topological sector $\kappa$ given as 
\begin{equation}
P_\kappa(x) = \vartheta_\kappa(z) \vartheta_\kappa^\dagger(z)
=\frac{1}{Z_\kappa}
\left (
\begin{array}{cc}
|z_1|^{2 \kappa} & z_1^\kappa {\bar z}_2^\kappa \\
z_2^\kappa {\bar z}_1^\kappa & |z_2|^{2 \kappa} \\
\end{array}
\right )
\label{eq:projection1}
\end{equation}
is a projector: $P_\kappa(x)^2 =P_\kappa(x)\,, P_\kappa(x)^\dagger = P_\kappa(x)$.

The field $n_a^\kappa(x)$ is associated to $P_\kappa(x)$ by the formulas
\begin{equation}
n_a^\kappa(x) = Tr \, \tau_a P_\kappa(x) = \vartheta_\kappa^\dagger(z) \tau_a \vartheta_\kappa (z) \,, \quad \quad
P_\kappa(x) = \frac{1 + \vec{\tau} \cdot \vec{n}^\kappa(x)}{2} \,.
\label{eq:fskappa}
\end{equation}
 
A phase change $z \rightarrow z e^{i \theta}$ induces the change $\vartheta_\kappa(z) \rightarrow \vartheta_\kappa(z) 
e^{i \kappa \theta}$. Nevertheless, this phase cancels in $\vartheta_\kappa(z) \vartheta_\kappa^\dagger(z)$ and $P_\kappa(x)$
is a function of $x$ only.

In \cite{bal} an intuitive argument as well as an explicit calculation is given to show that $\kappa$ appearing in equations
(\ref{eq:partial1}) through (\ref{eq:fskappa}) is indeed the winding number. Here we recollect the former. For 
$\kappa > 0$ consider the $\kappa$ points (up to an overall phase of $z$ which cancels out on $x$) of $S^2$ labeled by $\ell$:
\begin{equation}
z_\ell = ( z_1 e^{i \frac{2 \Pi}{\kappa} \ell} \,, z_2) \quad \quad \ell \in (0\,, \kappa - 1)\,.
\end{equation}
All $z_\ell$ map to the same point on the target manifold $S^2$ or equivalently, they all have the same projection via 
$P_\kappa(x)$, giving winding number $\kappa$.

It must be noticed that the form of $P_\kappa(x)$ is very particular. Nevertheless, the most general projector ${\cal P}_\kappa(x)$
can be obtained from 
\begin{equation}
{\cal P}_\kappa(x) = U(x) P_\kappa (x) U(x)^\dagger
\end{equation}
where $U(x) \in U(2)$ is a $2 \times 2$ unitary matrix. The field associated to  ${\cal P}_\kappa(x)$ is nothing but
\begin{equation}
n_a^{\kappa \prime}(x) = Tr \tau_a {\cal P}_\kappa(x) \,
\end{equation}
where $n_a^{\kappa \prime}(x) = R_{ab} n_b^{\kappa}(x)$, $U^\dagger \tau_a U = R_{ab} \tau_b$ and $R \in O(3)$.
The unitary transformation do not affect the the winding number since $\pi_2(U(2)) = \lbrace e \rbrace$.

\section{On the Actions}

A Euclidean action in the $\kappa$-th topological sector is given in terms of the fields $n^{\kappa}_a(x)$
\footnote{For brevity we drop the ``prime'' on the fields $n_a(x)$.} by
\begin{equation}
S_\kappa = - \frac{1}{8 \pi} \int_{S^2} d \Omega ({\cal L}_i n_a^\kappa) ({\cal L}_i n_a^\kappa)
\label{eq:action1}
\end{equation}
where ${\cal L}_i = -i (x \wedge \nabla)_i$ is the angular momentum operator and $d \Omega = d cos{\theta} 
d \psi$. In terms of the projectors, $S_\kappa$ can be expressed as 
\begin{equation}
S_\kappa = - \frac{1}{4 \pi} \int_{S^2} d \Omega \,  Tr \big ( {\cal L}_i  {\cal P}_\kappa \big) \,\big( {\cal L}_i  {\cal P}_\kappa \big ) \,.
\label{eq:action2}
\end{equation}
The well known formulae for the winding number and BPS bound of this model can also be rewritten in terms of the projectors
${\cal P}_\kappa$. The actions given in (\ref{eq:action1}) and (\ref{eq:action2}) both do have discrete versions when the 
${\mathbb C}P^1$ model is formulated on the fuzzy sphere $S_F^2$. However, it seems that the latter is better adapted for formulation 
of fuzzy ${\mathbb C}P^1$ sigma models; as will be discussed in section 6 it is possible to quantize the projectors in a 
straightforward manner. For a detailed discussion on the fuzzy ${\mathbb C}P^1$ model the reader is refered to \cite{bal}.

In section 5 we develop the supersymmetric extension of the projectors ${\cal P}_\kappa (x)$ and apply this result to 
the description of non-linear sigma model first on the supersphere and then on the fuzzy supersphere. The latter will require the
supersymmetric extension of quantized projectors.

\section{The Commutative and Non-Commutative (Fuzzy) Superspheres}

\subsection{The Supersphere $S^{(2,2)}$}

In this section we would like to collect some preliminary differential geometric and group theoretical formulae that is used to
characterize the supersphere $S^{2,2}$ and its non-commutative (fuzzy) version $S^{(2,2)}_F$. The details of the very brief 
discussion below can be found in \cite{peter, seckin}.

The structure underlying the supersphere $S^{(2,2)}$ comes from the Lie superalgebras $osp(2,1)$ and $osp(2,2)$ and their 
associated Lie supergroups $OSP(2,1)$ and $OSP(2,2)$. $osp(2,1)$ is build up of the Lie algebra $su(2)$ (even part) with generators
$L_i \,, (i = 1,2,3)$ and $su(2)$ spinors $V_\alpha (\alpha = +, -)$ (odd part). $osp(2,2)$ Lie superalgebra is constructed by 
augmenting $osp(2,1)$ generators with an additional pair of spinors $D_\alpha (\alpha = +, -)$ and an additional even generator 
$\Gamma$. The graded commutation relations of $osp(2,2)$ generators read 
\begin{eqnarray}
\lbrack L_i , L_j \rbrack &=& i \epsilon_{ijk} L_k \,, \quad   \lbrack L_i , V_\alpha \rbrack = \frac{1}{2} 
({\sigma}_i)_{ \beta \alpha} V_\beta \,, \quad \lbrace V_\alpha ,V_\beta \rbrace = \frac{1}{2} ( C \sigma_i)_{\alpha \beta} L_i \,, 
\nonumber \\
\lbrack L_i , \Gamma \rbrack &=& 0 \,, \quad  \lbrack \Gamma , V_\alpha \rbrack = D_\alpha \,, \quad  \lbrack \Gamma , D_\alpha 
\rbrack = V_\alpha \,, \quad \lbrack L_i , D_\alpha \rbrack = \frac{1}{2} ({\sigma}_i)_{ \beta \alpha} D_\beta \,, \nonumber \\
\lbrace D_\alpha ,D_\beta \rbrace &=& - \frac{1}{2} ( C \sigma_i)_{\alpha \beta} L_i \,, \quad
\lbrace D_\alpha ,V_\beta \rbrace = \frac{1}{4} C_{\alpha \beta} \Gamma \,.
\label{eq:generators}
\end{eqnarray}
where $i,j= 1,2,3$, $\alpha, \beta= \pm$ and $C = i \sigma_2$. 
The graded commutation relations for the $osp(2,1)$ generators is given by the first line of (\ref{eq:generators}).

In the corresponding enveloping algebras there are central polynomials - the Casimir operators in representations given by
the formulas:
\begin{eqnarray}
K_2^{osp(2,1)} &=& L_i L_i + C_{\alpha \beta} V_\alpha V_\beta \,, \nonumber \\
K_2^{osp(2,2)} &=& L_i L_i + C_{\alpha \beta} V_\alpha V_\beta - \Big( {C}_{\alpha \beta} D_\alpha D_\beta + \frac{1}{4} \Gamma^2 
\Big) \,. 
\label{eq:casimirs}
\end{eqnarray}

These Lie superalgebras are endowed with a grade dagger operation $\ddagger$ replacing the usual adjoint operation on the
Lie algebras. Generators of $osp(2,2)$ fulfill the following reality conditions implemented by $\ddagger$:
\begin{equation}
L_i^{\ddagger} = L_i^{\dagger} = L_i, \quad \quad
V_\alpha ^\ddagger = {C}_{\alpha \beta} V_\beta \,, \quad \quad D_\alpha ^\ddagger = - {C}_{\alpha \beta} D_\beta \,, \quad \quad   
\Gamma^{\ddagger} = \Gamma^{\dagger} = \Gamma \,.
\label{eq:reality}
\end{equation}
The reality conditions fulfilled by $osp(2,1)$ is obtained by restricting to the relations fulfilled by $L_i$ and $V_\alpha$. 
The graded conjugation is extended to homogeneous elements $A$ and $B$ in enveloping algebras by
\begin{equation}
(AB)^\ddagger = (-1)^{|A| |B|} B^\ddagger A^\ddagger \,.
\label{eq:envelopingex}
\end{equation}
Here $|A|$ and $|B|$ denote the degrees of $A$ and $B$, respectively. By linearity the conjugation is extended to the whole enveloping
algebra. The Casimir elements, given above, are real. 

The supersphere $S^{(2,2)}$ is the adjoint orbit of the Lie supergroup $OSP(2,1)$. It can be obtained through a super generalization
of the Hopf fibration for the 2-sphere. In the supersymmetric case this becomes $U(1) \rightarrow S^{(3,2)} \rightarrow S^{(2,2)}$   
where $S^{(3,2)} \equiv OSP(2,1)$ and 
\begin{equation}
S^{(2,2)} = S^{(3,2)} \diagup U(1) \,.
\end{equation}
The superspace ${\mathbb R}^{(3,2)}$ is defined as the algebra of polynomials in generators $x_i$ and $\theta_\alpha$ satisfying
reality conditions
\begin{equation}
x_i^\ddagger = x_i \,, \quad \quad \theta_\alpha^\ddagger = C_{\alpha \beta} \theta_\beta \,.
\end{equation}
These conditions are extended as in (\ref{eq:envelopingex}) to all polynomials.
The equation characterizing the adjoint orbit $S^{(2,2)}$ of $osp(2,1)$ is 
\begin{equation}
S^{(2,2)} =  \Big \langle (x_i \,, \theta_\alpha) \in {\mathbb R}^{(3,2)} \, 
\big | \, \, x_i^2 + C_{\alpha \beta} \, \theta_\alpha \theta_\beta = \frac{1}{4}\, \Big \rangle \,.
\end{equation} 

The action of $osp(2,1)$ on $S^{(2,2)}$ is the adjoint action and is given in terms of the differential operators
\begin{eqnarray} 
\ell_i &=& -i \varepsilon_{ijk} x_j \partial_k - \frac{1}{2}(\sigma_i)_{\beta \alpha} \theta_\beta \partial_{\theta^\alpha} \,, 
\nonumber \\
v_\alpha &=& - \frac{1}{2}(\sigma_i)_{\beta \alpha} \theta_\beta \partial_i + \frac{1}{2}(C \sigma_i)_{\alpha \beta} x_i
\partial_{\theta^\beta} \,.
\label{eq:diffop1}
\end{eqnarray}
corresponding to the $osp(2,1)$ generators $L_i$ and $V_\alpha$, respectively.
It can be extended to an $osp(2,2)$ action which is not an adjoint action but it is closely related to it. 
(for details see \cite{peter} \cite{seckin}). The additional differential operators have the form
\begin{eqnarray}
d_\alpha &=& -r \Big ( 1 + \frac{2}{r^2} \Big ) C_{\alpha \beta} \partial_{\theta^\beta} + \frac{1}{2r} (\sigma_i)_{\beta \alpha} 
\theta_\beta {\cal L}_i - \frac{\theta_\alpha}{2r} x_i \partial_i \,, \nonumber \\ 
\gamma &=& \Big ( \frac{\theta_+ x_3}{r} + \frac{\theta_- x_+}{r} \Big) \partial_+ +
\Big ( \frac{\theta_+ x_-}{r} - \frac{\theta_- x_3}{r} \Big) \equiv 2(\theta_- v_+ -\theta_+ v_-) \,.
\label{eq:diffop2}
\end{eqnarray}
corresponding to the generators $D_\alpha$ and $\Gamma$ of $osp(2,2)$. 

\subsection{The Fuzzy Supersphere $S_F^{(2,2)}$}

The fuzzy supersphere $S_F^{(2,2)}$ is obtained replacing $(x_i \,, \theta_\alpha) \in  {\mathbb R}^{(3,2)}$ 
by suitable rescaled $osp(2,1)$ generators
$X_i = \lambda L_i$ and $\Theta = \lambda V_\alpha$ with $\lambda$ determined by the value of $osp(2,1)$ Casimir operator:
\begin{equation}
\frac{1}{4 \lambda^{2}} = K_2^{osp(2,1)} \,.
\end{equation}
The fuzzy parameters then satisfy the supersphere's defining relation
\begin{equation}
X_i X_i + C_{\alpha \beta} \Theta_\alpha \Theta_\beta = \frac{1}{4} \,.
\label{eq:kazimir}
\end{equation}
The non-commutativity of the supersphere follows from the graded commutation relations of $X_i$ and $\Theta_\alpha$. For details
we refer the reader to \cite{peter}, \cite{seckin}.  

\section{Non-Linear Sigma Model on $S^{(2,2)}$}

\subsection{Preliminaries}

The superfield $\Phi$ on $S^{(2,2)}$ is a function of the variables $(x_i \,, \theta_\alpha)$; it is real provided that 
$\Phi^\ddagger = \Phi$. For a free real superfield multiplet the action is related to the $osp(2,1)$ invariant given 
as the difference of the quadratic Casimir operators:
\begin{equation}
K_2^{osp(2,1)} - K_2^{osp(2,2)} = {C}_{\alpha \beta} D_\alpha D_\beta + \frac{1}{4} \Gamma^2 \,.
\label{eq:casimir} 
\end{equation}
The action takes the form 
\begin{equation}
S^{SUSY} = \frac{1}{4 \pi} \int d \mu \Big ( d_\alpha \Phi d_\alpha \Phi + \frac{1}{2} \gamma \Phi  \frac{1}{2} \gamma \Phi \Big ) 
\label{eq:susyaction1}
\end{equation}
where $d \mu = d^3 x^i d \theta^+ d \theta^- \delta(x_i^2 + C_{\alpha \beta} \theta^\alpha \theta^\beta - \frac{1}{4})$,
and $d_\alpha$ and $\gamma$ are the differential operators given in (\ref{eq:diffop2}).

For a free triplet real superfield $\Phi^a = \Phi^a(x_i \,, \theta_\alpha)$, $(a= 1,2,3)$, we just replace in $\Phi$ by $\Phi^a$
(with the summation over repeated index $a$ understood). Now we define the $O(3)$ sigma model \cite{witten} by putting on $\Phi^a$ the
constraint
\begin{equation}
\Phi^a \Phi^a = 1 \quad \quad (a =1,2,3) \,.
\label{eq:constraint}
\end{equation}
Then (\ref{eq:susyaction1}) and (\ref{eq:constraint}) defines the non-linear sigma model on the supersphere $S^{(2,2)}$ 
with the target manifold being $S^2$.

The superfield $\Phi^a(x_i \,, \theta_\alpha)$ can be expanded in powers of $\theta_\alpha$ as
\begin{equation}
\Phi^a(x_i \,, \theta_\alpha) = n^a(x_i) + C_{\alpha \beta} \theta_\beta \psi^a_\alpha (x_i) + \frac{1}{2} F^a(x_i) 
C_{\alpha \beta} \theta_\alpha \theta_\beta
\label{eq:fieldexp}
\end{equation}
where $\psi^a(x_i)$ are two component Majorana spinors : $\psi_\alpha^{a \ddagger} = C_{\alpha \beta} \psi^a_\beta$, and $F^a(x_i)$
are auxiliary scalar fields. In terms of the component fields the constraint equation (\ref{eq:constraint}) splits to
\begin{equation}
n^a n^a = 1 \,, \quad \quad n^a F^a = \frac{1}{2} \psi^{a \ddagger} \psi^a \,, \quad \quad n^a \psi_\alpha^a = 0 \,.
\label{eq:compcons1}
\end{equation}

\subsection{Supersymmetric Extensions of Bott Projectors}

A possible supersymmetric extension of the projector ${\cal P}_\kappa(x)$  can be obtained in the following way.
Let ${\cal U}(x_i \,, \theta_\alpha)$ be a graded unitary operator with ${\cal U} {\cal U}^\ddagger = {\cal U}^\ddagger {\cal U} =1$. 
${\cal U}(x_i \,, \theta_\alpha)$ in general can be thought as a $2 \times 2$ supermatrix whose entries are functions on $S^{(2,2)}$.
${\cal U}(x_i \,, \theta_\alpha)$ acts on ${\cal P}_\kappa$ by conjugation and generates a set of supersymmetric extensions 
${\cal Q}_\kappa (x_i \,, \theta_\alpha)$:
\begin{equation}
{\cal Q}_\kappa(x_i \,, \theta_\alpha) = {\cal U}^\ddagger \, {\cal P}_\kappa(x) \, {\cal U} \,.
\label{eq:supertr1} 
\end{equation}
It is easy to see that ${\cal Q}_\kappa(x_i \,, \theta_\alpha)$ satisfies ${\cal Q}_\kappa^2(x_i \,, \theta_\alpha) = 
Q_\kappa(x_i \,, \theta_\alpha)$ and ${\cal Q}_\kappa^\ddagger(x_i\,, \theta_\alpha) = 
{\cal Q}_\kappa(x_i \,, \theta_\alpha)$. Thus ${\cal Q}_\kappa(x_i \,, \theta_\alpha)$ is a (super)-projector. 
The real superfield on $S^{(2,2)}$ associated to ${\cal Q}_\kappa(x_i \,, \theta_\alpha)$ is given by
\begin{equation}
\Phi_a^\prime (x_i \,, \theta_\alpha) = Tr \, \tau_a {\cal Q}_\kappa \,.
\label{eq:sf1}
\end{equation}
In order to perform a check that establishes that ${\cal Q}_\kappa(x_i \,, \theta_\alpha)$ are indeed the
supersymmetric projectors that reproduces the superfields on $S^{(2,2)}$ subject to 
\begin{equation}
\Phi^\prime_a \Phi^\prime_a = 1 \,,
\label{eq:31}
\end{equation}
we proceed as follows. First we expand ${\cal U}(x_i \,, \theta_\alpha)$ in powers of the Grassmann variables as 
\begin{equation}
{\cal U}(x_i \,, \theta_\alpha) = {\cal U}_0(x_i) + C_{\alpha \beta} \theta_\beta {\cal U}_\alpha(x_i) + \frac{1}{2} {\cal U}_2(x_i) 
C_{\alpha \beta} \theta_\alpha \theta_\beta 
\label{eq:expansion1}
\end{equation}
where ${\cal U}_0 \,, {\cal U}_\alpha (\alpha= \pm)$ and ${\cal U}_2$ are all $2 \times 2$ graded unitary matrices.
The requirement that ${\cal U}(x_i \,, \theta_\alpha)$ 
is graded unitary makes ${\cal U}_0(x_i)$ unitary, whereas ${\cal U}_\alpha(x_i)$ are uniquely determined by ${\cal U}_\alpha(x_i) 
= H_\alpha(x_i) {\cal U}_0(x_i)$ where $H_\alpha$ are $2 \times 2$ odd supermatrices with the reality condition 
$H_\alpha^\ddagger = - C_{\alpha \beta} H_\beta$. Moreover, with the ansatz that
${\cal U}_2 = A {\cal U}_0$ with $A$ being an arbitrary $2 \times 2$ even supermatrix, 
graded unitarity of ${\cal U}(x_i \,, \theta_\alpha)$ further restricts the symmetric part of $A$ as: 
\begin{equation}
A + A^\dagger = - C_{\alpha \beta} H_\alpha H_\beta \,.
\label{eq:constraint1}
\end{equation}    
Using the expansion (\ref{eq:expansion1}) in (\ref{eq:supertr1}) and subsequently the resulting expression in (\ref{eq:sf1})
together with the properties listed above it is straightforward to extract the component fields of the
superfield $\Phi_a^\prime(x_i \,, \theta_\alpha)$. We find 
\begin{eqnarray}
n_a^{\kappa \prime} &:=& Tr \, \tau_a U_0^\dagger {\cal P}_\kappa U_0 \,, \\   
\psi_\alpha^{a \prime} &:=& - 2 i (\vec{n}^{\kappa \prime} \times \vec{H}_\alpha^\prime)^a = Tr \, \tau_a U_0^\dagger 
\lbrack H_\alpha \,, {\cal P}_\kappa \rbrack U_0 \,,
\end{eqnarray}
and after using (\ref{eq:constraint1}) that
\begin{eqnarray}
F_a^\prime &:=& 4 ( \vec{H}_+^\prime \cdot \vec{H}_-^\prime) n_a^{\kappa \prime} - 2\vec{H}_+^{a \prime} (\vec{n}^{\kappa \prime} 
\cdot \vec{H}_-^\prime)
- (\vec{n}^{\kappa \prime} \cdot \vec{H}_+^\prime) 2\vec{H}_-^{a \prime} +i( \vec{n}^{\kappa \prime} \times (\vec{A}^\prime- \vec{A}^{\dagger \prime}))^a \\
&=& Tr \, \tau_a U_0^\dagger ( {\cal P}_\kappa A +A^\dagger {\cal P}_\kappa - C_{\alpha \beta} H_\beta {\cal P}_\kappa H_\alpha ) U_0 
\,. \nonumber 
\label{eq:fterm}
\end{eqnarray}  
where  $\vec{H}_\alpha^{a \prime}= H_\alpha^{a \prime} \tau^a$ and $\vec{A}^{a \prime} = A^{a \prime} \tau^a$.
By direct computation from above we find 
\begin{equation}
n_a^{\kappa \prime} n_a^{\kappa \prime} = 1 \,, \quad \quad n_a^{\kappa \prime} F_a^{\prime} 
= \frac{1}{2} \psi^{\ddagger \prime}_a \psi_a^\prime \,, \quad \quad
n_a^{\kappa \prime} \psi_\pm^{a \prime} = 0 \,.
\label{eq:constraints2}
\end{equation}
Comparing (\ref{eq:constraints2}) with (\ref{eq:compcons1}) we observe that they are identical. 
Therefore we conclude that the superfield associated
to the super-projector ${\cal Q}_\kappa$ is the same as the superfield in supersymetric non-linear sigma model of the previous 
subsection.
 
\subsection{SUSY Action Revisited}

We are now ready to give the formulation of non-linear sigma model on the supersphere using the (super)-projectors. In close analogy
with the ${\mathbb C}P^1$ case the supersymmetric action in (\ref{eq:susyaction1}) with the constraint (\ref{eq:constraint}) 
translates to
\begin{equation}
S_\kappa^{SUSY} = \frac{1}{2 \pi} \int d \mu \, Tr \Big \lbrack (d_\alpha {\cal Q}_\kappa) (d_\alpha {\cal Q}_\kappa) + \frac{1}{4} 
(\gamma {\cal Q}_\kappa) (\gamma {\cal Q}_\kappa) \Big \rbrack \,.
\label{eq:superaction2}
\end{equation} 
The even part of this action, as well as the one given in (\ref{eq:susyaction1}) is nothing but the action $S_\kappa$ of the 
${\mathbb C}P^1$ theory given in (\ref{eq:action2}) and  (\ref{eq:action1}), respectively. In other words, the action 
$S_\kappa^{SUSY}$ is the supersymmetric extension of $S_\kappa$ on $S^2$ to $S^{(2,2)}$. Thus in the supersymmetric theory
it is possible to interpret the index $\kappa$ carried by the action as the winding number of the corresponding ${\mathbb C}P^1$
theory.
    
We recall that $d_\alpha$ and $\gamma$ are both derivations in the superalgebra $Osp(2,2)$. Therefore they obey a graded Leibnitz rule
and from ${\cal Q}_\kappa^2 = {\cal Q}_\kappa$ we find
\begin{equation}
{\cal Q}_\kappa d_\alpha {\cal Q}_\kappa = d_\alpha {\cal Q}_\kappa ({\bf 1} -{\cal Q}_\kappa) \,.
\label{eq:property1}
\end{equation}
This enables us to write
\begin{equation}
Tr d_\alpha {\cal Q}_\kappa ({\bf 1} - {\cal Q}_\kappa) d_\alpha {\cal Q}_\kappa = Tr ({\bf 1} -{\cal Q}_\kappa)
(d_\alpha {\cal Q}_\kappa)^2 = \frac{1}{2} Tr (d_\alpha {\cal Q}_\kappa)^2 \,.
\label{eq:property2}
\end{equation}
Equations (\ref{eq:property1}) and (\ref{eq:property2}) continue to hold when $d_\alpha$ is replaced by $\gamma$ as well. The action
then takes the form 
\begin{equation}
S_\kappa^{SUSY} = \frac{1}{\pi} \int d \mu\,  Tr \Big \lbrack {\cal Q}_\kappa (d_\alpha {\cal Q}_\kappa) (d_\alpha {\cal Q}_\kappa) 
+ \frac{1}{4} {\cal Q}_\kappa (\gamma {\cal Q}_\kappa) (\gamma {\cal Q}_\kappa) \Big \rbrack \,.
\label{eq:action3}
\end{equation}  
It is possible that this form of the action could play an important role for obtaining an supersymmetric generalization of the BPS
equation since an analogues expression in the ${\mathbb C}P^1$ case \cite{bal} have been employed to achieve this result.

\section{Fuzzy Projectors and Sigma Models}

\subsection{Fuzzy ${\mathbb C}P^1$ Model}

In \cite{bal} the ${\mathbb C}P^1$ model has been quantized as follows. Let $\xi = (\xi_1 \,, \xi_2) \in {\mathbb C}^2 \diagdown \{0\}$. In terms of
$\xi$ we define
\begin{equation}
z = \frac{\xi}{|\xi|} \,, \quad \quad |\xi| = \sqrt{|\xi_1|^2 + |\xi_2|^2} \,, \quad \quad x_i = z^\dagger x_i z \,.
\end{equation}
$\xi_\alpha$ and ${\bar \xi}_\alpha$ are quantized by replacing them with a pair of annihilation $a_\alpha$ and creation 
$a_\alpha^\dagger$ operators respectively. With this substitution $|\xi|$ becomes the square root of the number operator and we have 
\begin{eqnarray}
{\hat N} &=& {\hat N}_1 + {\hat N}_2 \,, \quad {\hat N}_1 = a_1^\dagger a_1 \,, \quad N_2 = a_2^\dagger a_2 \nonumber \\
{\hat z}_\alpha^\dagger &=& \frac{1}{\sqrt{{\hat N}}} a_\alpha^\dagger = a_\alpha^\dagger \frac{1}{\sqrt{{\hat N}+1}} \,, \quad 
{\hat z}_\alpha = \frac{1}{\sqrt{{\hat N}+1}} a_\alpha = a_\alpha \frac{1}{\sqrt{{\hat N}}} \,, \nonumber \\
{\hat x}_i &=& \frac{1}{{\hat N}} a^\dagger \tau_i a \,.
\end{eqnarray}

In the light of this conjecture it is easy to see that the quantized version of the partial isometry $\vartheta_\kappa^\dagger$ 
defined in (\ref{eq:partial1}) and its Hermitian conjugate reads
\begin{eqnarray}
{\hat \vartheta}_\kappa^\dagger &=& \frac{1}{\sqrt{{\hat Z}_\kappa}}
\left (
\begin{array}{cc}
a_1^{\dagger \kappa} & a_2^{\dagger \kappa} \\
\end{array}
\right ) \,, \quad \quad
{\hat \vartheta}_\kappa =
\left (
\begin{array}{c}
a_1^\kappa \\
a_2^\kappa \\
\end{array}
\right ) 
\frac{1}{\sqrt{{\hat Z}_\kappa}} \,, \quad \quad {\hat \vartheta}_\kappa^\dagger{\hat \vartheta}_\kappa = {\bf 1} \,, \\
{\hat Z}_\kappa &=& {\hat Z}_\kappa^{(1)} + {\hat Z}_\kappa^{(2)} \,, \quad  {\hat Z}_\kappa^{(\alpha)} = {\hat N}_\alpha 
({\hat N}_\alpha -1) \ldots ({\hat N}_\alpha - \kappa + 1) \nonumber \,. 
\end{eqnarray}
The fuzzy analogue of (\ref{eq:projection1}) can now be written as
\begin{equation}
{\hat P}_\kappa(x) = {\hat \vartheta}_\kappa {\hat \vartheta}_\kappa^\dagger =
\left (
\begin{array}{cc}
a_1^{\kappa} \frac{1}{{\hat Z}_\kappa} a_1^{\dagger \kappa} & a_1^{\kappa} \frac{1}{{\hat Z}_\kappa} a_2^{\dagger \kappa}  \\
a_2^{\kappa} \frac{1}{{\hat Z}_\kappa} a_1^{\dagger \kappa} & a_2^{\kappa} \frac{1}{{\hat Z}_\kappa} a_2^{\dagger \kappa}  \\
\end{array}
\right )
\label{eq:projection2}
\end{equation}
where for example
\begin{equation}
a_1^\kappa \frac{1}{{\hat Z}_\kappa} = \frac{1}{({\hat N}_1 + \kappa) \hdots ({\hat N}_1 + 1) + {\hat Z}_\kappa^{(2)}} a_1^\kappa
\,, \quad a_1^\kappa a_1^{\dagger \kappa} = ({\hat N}_1 + \kappa) \hdots ({\hat N}_1 + 1) \,.
\end{equation}

The unitary matrix $U$ introduced to generate all possible projectors ${\cal P}_\kappa$ from $P_\kappa$ also 
have fuzzy analogue.
It is a $2 \times 2$ unitary matrix ${\hat U}$, with matrix entries being polynomials in $a_\alpha^\dagger a_\beta$. Thus the most
general fuzzy projectors are
\begin{equation}
{\hat {\cal P}}_\kappa = {\hat U} {\hat P}_\kappa {\hat U}^\dagger \,.
\end{equation}    

From (\ref{eq:projection2}) it is clear that ${\hat {\cal P}}_\kappa$ acts in general on $ {\cal F}^2:= {\cal F} \otimes 
{\mathbb C}^2$ where ${\cal F}$ stands for the standard Fock space. It also follows from (\ref{eq:projection2}) that ${\hat P}_\kappa$
commutes with the number operator ${\hat N}$, as can be checked directly. Consequently, we can restrict ourselves to work on a finite 
dimensional subspace ${\cal F}_n$ of dimension $n+1$ of ${\cal F}$. Then ${\hat {\cal P}}_\kappa$ act on the finite dimensional 
Hilbert space ${\cal F}_n^2 := {\cal F}_n \otimes {\mathbb C}^2$ and this allows one to formulate a finite dimensional matrix 
model for projectors ${\hat {\cal P}}_\kappa$.

In \cite{bal} this has been done to write down the fuzzy ${\mathbb C}P^1$ model. They found that the fuzzy action corresponding 
to (\ref{eq:action2}) is
\begin{equation}
S_{F \,, \kappa} =  \frac{1}{4 \pi} \frac{1}{2 (n+1)} Tr_{{\hat N} = n} ({\cal L}_i  {\widehat {\cal P}}_\kappa) ({\cal L}_i  {\widehat {\cal P}}_\kappa)
\label{eq:fuzzyaction1}
\end{equation} 
where ${\cal L}_i  {\widehat {\cal P}}_\kappa = \lbrack L_i \,, {\widehat {\cal P}}_\kappa \rbrack$ and the trace is over ${\cal F}_n^{(2)}$.

\subsection{Fuzzy Supersymmetric Model}

In much the same way the supersymmetric projectors ${\cal Q}_\kappa$ have been constructed from ${\cal P}_\kappa$ in section 5, we 
can construct supersymmetric extensions of ${\widehat {\cal P}}_\kappa$ by the graded unitary transformation
\begin{equation}
{\widehat {\cal Q}}_\kappa = {\hat {\cal U}}^\ddagger {\widehat {\cal P}}_\kappa {\hat {\cal U}}
\label{eq:fuzzyp}
\end{equation}
where ${\hat {\cal U}}$ is a $2 \times 2$ supermatrix whose entries are polynomials in $a_\alpha^\dagger a_\beta$ and 
$b^\dagger b$ and where $b$ and $b^\dagger$ are fermionic annihilation and creation operators with the anti-commutation relation 
$\{b \,, b^\dagger \} =1$.
 
${\widehat {\cal Q}}_\kappa$ defined (\ref{eq:fuzzyp}) acts on the finite dimensional space ${\tilde {\cal F}}^2_n = 
{\tilde {\cal F}}_n \otimes {\mathbb C}^2$. Here ${\tilde {\cal F}}$ is the $n+1$ dimensional subspace of the Hilbert 
space ${\tilde {\cal F}}$ spanned by the kets $\big | n_1, n_2, n_3 \rangle$ where $n_3$ labels the fermionic part taking on the values 
$0$ and $1$ only. It is readily seen that ${\widehat {\cal Q}}_\kappa$ commutes with the supersymmetric number operator 
${\widehat {\cal N}} = a_\alpha^\dagger a_\alpha + b^\dagger b$. 
In close analogy with the fuzzy  ${\mathbb C}P^1$ model, it is now possible to write down a finite dimensional (super)matrix model
for the  (super)-projectors ${\widehat {\cal Q}}_\kappa$.

Making use of (\ref{eq:casimir}) once more the action for the fuzzy supersymmetric model is given as
\begin{equation}
S_{F \,, \kappa}^{SUSY} = \frac{1}{2\pi} \, Str_{{\widehat {\cal N}} =n}  \, \Big \lbrack ({\cal D}_\alpha {\widehat {\cal Q}}_\kappa) 
({\cal D}_\alpha {\widehat {\cal Q}}_\kappa) + \frac{1}{4} (\Xi {\widehat {\cal Q}}_\kappa) (\Xi {\widehat {\cal Q}}_\kappa) \Big \rbrack 
\,,
\label{eq:superaction2f}
\end{equation}
where ${\cal D}_\alpha {\widehat {\cal Q}}_\kappa = \lbrace D_\alpha \,, {\widehat {\cal Q}}_\kappa \rbrace$ and $\Xi {\widehat {\cal Q}}_\kappa
=\lbrack \Gamma \,, {\widehat {\cal Q}}_\kappa \rbrack$. ``Str'' in the above expression is the supertrace over the finite
dimensional space ${\tilde {\cal F}}^2_n$. Obviously, in the large  ${\widehat {\cal N}} = n$ limit (\ref{eq:superaction2f})
approximates the action given in (\ref{eq:superaction2}).

\section{Conclusions}

In this paper we have obtained the fuzzy version of supersymmetric non-linear sigma model on $S^{(2,2)}$. Our approach has utilized
the use of supersymmetric extensions of the Bott projectors and generalized results of ${\mathbb C}P^1$ model to supersymmetric
theories. A natural question to be addressed is the supersymmetric generalization of the BPS equation. We hope to report any 
development on this issue elsewhere.
\newline
\vskip 1em
{\bf Acknowledgments}
\newline
I would like to thank A.P. Balachandran for his supervision through the course of this study. I also would like to thank
Peter  Pre\v{s}najder for his extensive reading of the manuscript and critical comments and suggestions.
This work has been supported in part by DOE and NSF under the contract numbers DE-FG02-85ER40231 and INT9908763 respectively.


\vskip 1em

\begin{thebibliography}{10}

\bibitem{bal} A.~P.~Balachandran and G.~Immirzi, ``Fuzzy Nambu-Goldstone physics,'' Int. J. Mod. Phys. {\bf A 18} (2003) 5981,
arXiv:hep-th/0212133.

\bibitem{paper1}
J.~Madore,
``The fuzzy sphere,''
Class.\ Quant.\ Grav.\  {\bf 9}, 69 (1992);

\bibitem{paper2}
H.~Grosse, C.~Klimcik and P.~Presnajder,
``Towards finite quantum field theory in noncommutative geometry,''
Int.\ J.\ Theor.\ Phys.\  {\bf 35}, 231 (1996)
[arXiv:hep-th/9505175].

\bibitem{paper3}
P.~Presnajder,
``The origin of chiral anomaly and the noncommutative geometry,''
J.\ Math.\ Phys.\  {\bf 41}, 2789 (2000)
[arXiv:hep-th/9912050];

A.~P.~Balachandran, T.~R.~Govindarajan and B.~Ydri,
``The fermion doubling problem and noncommutative geometry,''
Mod.\ Phys.\ Lett.\ A {\bf 15}, 1279 (2000)
[arXiv:hep-th/9911087];

A.~P.~Balachandran and S.~Vaidya,
``Instantons and chiral anomaly in fuzzy physics,''
Int.\ J.\ Mod.\ Phys.\ A {\bf 16}, 17 (2001);

A.~P.~Balachandran and G.~Immirzi,
``The fuzzy Ginsparg-Wilson algebra: A solution of the fermion doubling  problem,''
Phys.\ Rev.\ D {\bf 68}, 065023 (2003)
[arXiv:hep-th/0301242];

B.~Ydri,
``Noncommutative chiral anomaly and the Dirac-Ginsparg-Wilson operator,''
JHEP {\bf 0308}, 046 (2003)
[arXiv:hep-th/0211209].

\bibitem{paper4}
S.~Vaidya,
``Perturbative dynamics on fuzzy S(2) and RP(2),''
Phys.\ Lett.\ B {\bf 512}, 403 (2001)
[arXiv:hep-th/0102212];

C.~S.~Chu, J.~Madore and H.~Steinacker,
``Scaling limits of the fuzzy sphere at one loop,''
JHEP {\bf 0108}, 038 (2001) [arXiv:hep-th/0106205].

B.~P.~Dolan, D.~O'Connor and P.~Presnajder,
``Matrix phi**4 models on the fuzzy sphere and their continuum limits,''
JHEP {\bf 0203}, 013 (2002)
[arXiv:hep-th/0109084];

\bibitem{paper5}
A.~P.~Balachandran and S.~Kurkcuoglu,
``Topology change for fuzzy physics: Fuzzy spaces as Hopf algebras,''
arXiv:hep-th/0310026.

\bibitem{bal1}
S.~Baez, A.~P.~Balachandran, B.~Idri and S.~Vaidya,
``Monopoles and solitons in fuzzy physics,''
Commun.\ Math.\ Phys.\  {\bf 208}, 787 (2000)
[arXiv:hep-th/9811169].

\bibitem{trg} T.~R.~Govindarajan and E.~Harikumar, ``O(3) sigma model with Hopf term on fuzzy sphere,''
Nucl.\ Phys.\ B {\bf 655}, 300 (2003) [arXiv:hep-th/0211258].

\bibitem{wegge-olsen} N.E. Wegge Olsen, {\it K-theory and $C^*$-Algebras-a Friendly Approach}, Oxford University Press, Oxford, 1993.

\bibitem{peter}
H.~Grosse, C.~Klimcik and P.~Presnajder,
``Field theory on a supersymmetric lattice,''
Commun.\ Math.\ Phys.\  {\bf 185}, 155 (1997)
[arXiv:hep-th/9507074];

H.~Grosse, C.~Klimcik and P.~Presnajder,
``Topologically nontrivial field configurations in noncommutative geometry,''
Commun.\ Math.\ Phys.\  {\bf 178}, 507 (1996)
[arXiv:hep-th/9510083];

C.~Klimcik,
``An extended fuzzy supersphere and twisted chiral superfields,''
Commun.\ Math.\ Phys.\  {\bf 206}, 587 (1999)
[arXiv:hep-th/9903202].

\bibitem{seckin} A.~P.~Balachandran, S.~Kurkcuoglu and E.~Rojas, ``The star product on the fuzzy supersphere,''
JHEP {\bf 0207}, 056 (2002) [arXiv:hep-th/0204170].

 
\bibitem{witten} 
P.~Di Vecchia and S.~Ferrara,
``Classical Solutions In Two-Dimensional Supersymmetric Field Theories,''
Nucl.\ Phys.\ B {\bf 130}, 93 (1977);

E.~Witten, ``A Supersymmetric Form Of The Nonlinear Sigma Model In Two-Dimensions,''
Phys.\ Rev.\ D {\bf 16}, 2991 (1977).

\end{thebibliography}
\end{document}